\documentclass[12pt]{article}
\usepackage{mathrsfs}
\usepackage{amssymb}
\usepackage{amsmath}
\usepackage[amsmath,thmmarks]{ntheorem}
\usepackage{fullpage}

\usepackage{graphicx}
\usepackage{bm}
\usepackage{slashed}
\usepackage{color}

\def\lsim{\mathrel{\rlap{\lower4pt\hbox{\hskip1pt$\sim$}}
    \raise1pt\hbox{$<$}}}                
\def\gsim{\mathrel{\rlap{\lower4pt\hbox{\hskip1pt$\sim$}}
    \raise1pt\hbox{$>$}}}                

\def\qslash{q\!\!\!\slash}

\def\OMIT#1{}

\newcommand{\be}{\begin{eqnarray}}
\newcommand{\ee}{\end{eqnarray}}

\newcommand{\nn}{\nonumber}
\newcommand{\w}{\omega}

\newcommand{\bea}{\begin{eqnarray}}
\newcommand{\eea}{\end{eqnarray}}

\def\lsim{\mathrel{\rlap{\lower4pt\hbox{\hskip1pt$\sim$}}
    \raise1pt\hbox{$<$}}}                
\def\gsim{\mathrel{\rlap{\lower4pt\hbox{\hskip1pt$\sim$}}
    \raise1pt\hbox{$>$}}}                

\def\qslash{q\!\!\!\slash}

\def\OMIT#1{}

\begin{document}
\ifpdf
\DeclareGraphicsExtensions{.pdf, .jpg,.ps,.eps}
\newcommand{\picspace}{\vspace{-2.5in}}
\newcommand{\picspacehalf}{\vspace{-1.75in}}
\else
\DeclareGraphicsExtensions{.eps, .jpg,.ps}
\newcommand{\picspace}{\vspace{0in}}
\newcommand{\picspacehalf}{\vspace{0in}}
\fi

\begin{titlepage}

\makebox[6.5in][r]{\hfill ANL-HEP-PR-12-28}

\vskip1.40cm
\begin{center}
{\Large {\bf Gauge-boson production with multiple jets near threshold}} \vskip.5cm
\end{center}
\vskip0.2cm

\begin{center}
{\bf Xiaohui Liu, Sonny Mantry,  Frank Petriello}
\end{center}
\vskip 6pt
\begin{center}
{\it Department of Physics \& Astronomy, Northwestern University, Evanston, IL 60208, USA} \\
{\it High Energy Physics Division, Argonne National Laboratory, Argonne, IL 60439, USA} \\
\end{center}

\vglue 0.3truecm

\begin{abstract}
\vskip 3pt \noindent

Signatures of new physics beyond the Standard Model are often characterized by large missing transverse energy ($\not\!\! E_T$) produced in association with multiple jets. The dominant Standard Model background to such processes comes from gauge-boson production in association with jets. A standard search strategy involves looking for an excess in the $m_{eff}$ distribution, where $m_{eff}=  \not\!\! E_T +\sum_{J} p^T_J$ and $p^T_J$ denotes the transverse momentum of the $J$-th jet. The region of large $m_{eff}$  is dominated by jet production  near threshold, giving rise to large Sudakov logarithms that can change the magnitude and shape of the $m_{eff}$ distribution. We present an effective theory framework for the resummation of such  threshold logarithms. We perform an analysis for exclusive jet production using the N-jettiness global event shape, which allows theoretical control to also be maintained over large logarithms induced by vetoing additional jets.  As a first step, we give explicit numerical results with next-to-leading-log (NLL) resummation for $pp \to \gamma + 2$ jets in the large $m_{eff}$ region.

\end{abstract}

\end{titlepage}
\newpage

\section{Introduction}

The production of gauge bosons in association with jets is a dangerous background to searches for new physics at the LHC.  When the gauge boson in question is a $Z$-boson that decays 
into neutrinos, the resulting missing energy plus multi-jet final state is a dominant background to searches for supersymmetry (SUSY) and other models with dark-matter candidates, such as Little Higgs with T-parity or universal extra dimensions.  Searches for SUSY and these other theories in the multi-jet plus missing energy channel typically utilize a shape 
difference between signal and background in the tail of the $m_{eff}$ distribution, where $m_{eff}$ is defined as the scalar sum of the missing transverse energies and jet transverse momenta:
\begin{equation}
\label{meff}
m_{eff} = \not\!\! E_T +\sum_J p^T_J.
\end{equation}
We do not distinguish between the transverse energy and momentum of a jet.  The sum over jet transverse momenta may be taken either over only the leading 2, 3 or 4 jets, or may be taken over all identified jets in the event; ATLAS~\cite{Aad:2011ib} follows the first of these approaches, while CMS utilizes the second~\cite{Khachatryan:2011tk}.  Considering different jet-multiplicity bins may help probe non-standard SUSY theories, such as those with 
compressed spectra~\cite{LeCompte:2011cn}.

Motivated by its importance, numerous theoretical efforts have been devoted to calculating precisely gauge bosons produced in association with jets in the Standard Model.  Recent advances have allowed $W,Z+3$ jets~\cite{Ellis:2009zw,Berger:2009zg,Berger:2010vm} and even $W,Z+4$ jets~\cite{Berger:2010zx,Ita:2011wn} to be calculated to next-to-leading order (NLO) accuracy in perturbative QCD.  These calculations 
substantially reduce the residual scale uncertainty of the prediction.  However, heavy supersymmetric states will populate the tail of the $m_{eff}$ distributions, where fixed-order perturbation theory at NLO may miss important corrections appearing at higher orders.  In particular, large logarithms of the schematic form ${\rm ln} (1-m_{eff}^2/s)$, with $s$ the center-of-mass energy squared of the hadronic collisions, 
may be induced by the implicit restriction on soft gluons that comes from having nearly all the energy go into the leading few jets and $\not\!\! E_T$.  Such effects can be exacerbated for gluonic initial states, due to the steeply falling gluon distribution at high Bjorken $x$.  They would increase for high $m_{eff}$, potentially mimicking a SUSY signal.  The resummation of such soft-gluon logarithms following the approach of Refs.~\cite{Sterman:1986aj,Catani:1989ne} has been previously studied for the related single-inclusive jet production $p_T$ distribution~\cite{Kidonakis:2000gi}, and numerical results for the Tevatron and RHIC were presented in Ref.~\cite{deFlorian:2007fv}.  The resummation effects were found to be moderate but increasing at high $p_T$, generating the aforementioned shape difference.  In the 2011 data set, both ATLAS and CMS saw events with $m_{eff}$ near 2 TeV, and with the higher-luminosity run of 2012, events even closer to machine threshold will be observed.  A study of threshold resummation of gauge boson plus multi-jet processes at the LHC is therefore 
warranted.  

We begin a study of large logarithmic corrections to gauge boson plus multi-jet production in this manuscript by considering the next-to-leading logarithmic (NLL) threshold resummation for 
\bea
pp\to\gamma+2\text{ jets}
\eea
at the LHC.  This process represents a first step toward a study of the 
$W/Z+n$ jet process, but is also interesting on its own as a possible calibration process for missing energy plus jet backgrounds~\cite{Bern:2011pa}.  We utilize the Soft-Collinear Effective Theory (SCET)~\cite{Bauer:2000ew,Bauer:2000yr,Bauer:2001yt,Bauer:2002nz} to study the effect of threshold logarithms.  A formulation of threshold resummation within SCET applicable to multi-jet processes was given 
in Ref.~\cite{Bauer:2010vu}, and we build on the approach outlined there.  In addition we incorporate N-jettiness~\cite{Stewart:2010tn}, an event-shape based formalism for exclusive jet production, in our analysis. The N-jettiness formalism allows one to veto additional jets and maintain theoretical control of the induced logarithmic corrections to all orders in perturbation theory.  All jet-algorithm dependence is power suppressed. Such an event-shape based analysis simplifies higher-order calculations, since these do not depend on the jet algorithm up to power corrections. The production of electroweak gauge bosons at high transverse momentum, which is  inclusive in the recoiling hadronic radiation and consequently does not a require jet definition, has been previously studied~\cite{Becher:2009th} using SCET.  

We derive a factorization formalism for the production of a gauge boson in association with N-jets in the threshold limit that takes the schematic form
\bea
\label{facschem}
{d\sigma} \sim H\otimes J_1\otimes \cdots \otimes J_N \otimes S \otimes f\otimes f.
\eea
Here $H$ is the hard function that encodes the physics of the production at the hard scale and can be obtained from known fixed-order calculations, $J_i$ denote jet functions which encode the effects of collinear radiation within the $N$-jets in the final state, $S$ describes soft radiation inside and outside the N-jets throughout the event, and $f$ denotes the standard parton distribution function.  
The cross-section in Eq.~(\ref{facschem}) is inclusive in $\>  \not\!\! E_T $ and the transverse jet momenta $p_{J}^T$ subject to the constraint of Eq.~(\ref{meff}).  The phase space in the region of large $m_{eff}$ is then dominated by configurations corresponding to $N$ hard jets with only soft radiation occupying the region between the beam and jet directions.  We discuss this point  later in the text. In the threshold region,  factorization is given in terms of the initial state PDFs and does not involve beam functions~\cite{Fleming:2006cd} that can arise in other processes away from threshold where energetic collinear radiation is allowed in the beam directions.

Another potentially large logarithmic correction can arise when the gauge boson is either soft or collinear to either the initial state or one of the final-state jets.  However, experimental analyses often demand that the ratio $\not\!\! E_T / m_{eff}$ be greater than some minimum value, and impose a minimum angular separation between the missing-energy vector and each jet~\cite{LeCompte:2011cn}.  These constraints reduce the effect of such terms, and we consequently do not consider their effect here. 

We present numerical results for the 7 TeV LHC 
production of $\gamma+2$ jets, and in particular study the enhancement of the NLL cross section over the leading-order (LO) result.  The ratio of the NLL result over the LO one is approximately 1.5 at $m_{eff} \approx 2$ TeV, and increases further closer to machine threshold.  This shape difference demonstrates the importance of having threshold logarithms under theoretical control in searches at the LHC.

Our paper is organized as follows.  We review the kinematics relevant to the study of $\gamma+2$ jet production in Section~\ref{sec:kinematics}.  We show here that the phase-space region with a large separation between jet transverse momenta gives a suppressed contribution to the high $m_{eff}$ distribution.  In Section~\ref{sec:scetop} the operator basis 
we use to match QCD onto SCET is introduced.  We discuss the factorization theorem for the threshold region in Section~\ref{sec:fac}.  Numerical results are presented in Section~\ref{sec:num}.  Our formalism confers theoretical control over logarithms associated with vetoing extra jets in addition to threshold logarithms.  We discuss potential applications of this feature, as well as open questions and 
other future directions, in Section~\ref{sec:conc}.  A detailed 
description of the hard-matching coefficients and the solutions to the renormalization-group evolutions equations for the hard, soft, and jet functions, are given in the Appendix.

\section{Kinematics}
\label{sec:kinematics}

In this section we describe the kinematics and relevant degrees of freedom for the production of N-jets together with a gauge boson near threshold. We also elaborate on the definition of the threshold region which dominates the phase space in the large $m_{eff}$ region. We closely follow the work of Ref.~\cite{Bauer:2010vu} and adapt their formalism to an event-shape analysis using N-jettiness~\cite{Stewart:2010tn} instead of traditional  jet algorithms. 

Each jet is characterized by its transverse momentum $p_J^T$ and pseudorapidity $\eta_J$.   We demand a minimum partonic center of mass energy $\hat{s}_{min}$ defined as
\bea
\label{smin}
\hat{s}_{min} = (q+\sum_{i=1}^N p_J^i)^2,
\eea
where the massless momenta $p_J^i$ are defined as
\bea
\label{pj}
p_J\equiv (p_J^T\> \text{cosh}\> \eta_J,\bold{p}_J^T, p_J^T\> \text{sinh}\> \eta_J)
\eea
and $q$ is the gauge-boson momentum.  The ratios between $\hat{s}_{min}$ and the partonic and hadronic center-of-mass energies squared $\hat{s}$ and $s$ are characterized by the variables $z$ and $\tau$:
\bea
z=\frac{\hat{s}_{min}}{\hat{s}}, \qquad \tau= \frac{\hat{s}_{min}}{s}, \qquad \tau \leq z \leq 1.
\eea
The limit of hadronic threshold $\tau\to 1$ automatically forces the partonic threshold $z\to 1$. However, other dynamical effects such as the steepness of parton luminosities can force the partonic threshold limit $z\to 1$ even away from hadronic threshold~\cite{Appell:1988ie,Becher:2007ty}.

In the limit of partonic threshold, the initial-state collinear partons can emit only soft gluons. Additional collinear radiation arises from emissions off the final-state hard partons that form the jets. Thus, the relevant degrees of freedom in the limit of partonic threshold correspond to soft and collinear modes with momentum scalings
\bea
\label{scaling}
\text{collinear:} \>p_c \sim \sqrt{\hat{s}} (\lambda^2,1,\lambda), \qquad \text{soft:} \>k_s\sim \sqrt{\hat{s}}(\lambda^2,\lambda^2,\lambda^2),
\eea
where $\lambda \sim \sqrt{1-z}$. The momenta are decomposed in terms of light-cone coordinates
as $p=(p^+,p^-,p_\perp)$.  The collinear modes along each jet are decomposed in terms of light-cone coordinates with the spatial components of the light-cone vector aligned along the jet direction.  Momentum conservation at the partonic level is given by
\bea
p_I=q + k_s + \sum_i^N p^c_i,
\eea
where $p_I$ denotes the total initial state partonic momentum so that $p_I^2=\hat{s}$. The total final-state partonic soft momentum is denoted by $k_s$, while $p_c^i$ denotes the momentum of the hard parton that eventually forms the $i$-th final-state jet.

To define N-jettiness~\cite{Stewart:2010tn}, the final-state particles are grouped into regions that are associated with either one of the jets or the beam directions using a well-defined distance measure.  The total momentum of a jet then corresponds to the sum of the momenta of all the particles grouped into the corresponding jet region. The total N-jettiness $\tau_N$ for the event is defined as
\bea
\label{tauN}
\tau_N = \sum_i^{N+2}\tau_N^i = \sum_i 2 \hat{q}_i\cdot P_i, \qquad P_i^\mu=\sum_k p_k^\mu \>\>\Pi_{_{j\neq i}}\> \theta(\hat{q}_j\cdot p_k - \hat{q}_i\cdot p_k).
\eea
The $\hat{q}_i$ denote massless reference vectors along the $N$ jet and two beam directions, $P_i$ is the total momentum of all particles grouped into region $i$, and $\tau_N^i$ is the contribution of region $i$ to the total jettiness $\tau_N$. The total momentum of the $i$-th jet is defined to be $P_i$, corresponding to the total momentum of all particles in the $i$-th region.  One can also derive cross-sections that are differential in the $\tau_N^i$ of the various jet and beam regions.

All the collinear particles associated with the $i$-th jet direction are naturally grouped into the $i$-th region. The total final-state soft momentum can be decomposed as
\bea
\label{softmom}
k_s= k_{out} + \sum_i^N k_i,
\eea
where $k_i$ is the momentum contribution of the soft radiation to the $i$-th jet and $k_{out}$ is the total soft momentum that is not grouped with any of the jets. In other words, the momentum  $k_{out}$ is defined as the total momentum of soft particles that are grouped with one of the two beam directions.

It is useful to decompose the total final-state momentum into two parts. The first part corresponds to the minimum momentum needed to create the color-neutral sector with momentum $q^\mu$ and N-jets with massless momenta $p_J$ as in Eq.~(\ref{pj}). This minimum total momentum leads to a partonic center-of-mass energy squared given by $\hat{s}_{min}$ of Eq.~(\ref{smin}). The remaining part of the total momentum leads to the actual partonic center-of-mass energy squared $\hat{s} \geq \hat{s}_{min}$.  This decomposition can be made explicit  by noting that
any four momentum $p^\mu$ can be written in terms of the massless momentum $p_J$ of Eq.~(\ref{pj}) as
\bea
p^\mu= p_J^\mu + p^+ v^\mu,
\eea
where $v^\mu=(1,\bold{0})$ and $p^+=p^0-|\bold{p}|$. This follows from $p^0=p_J^0+(p^0-p_J^0)$ and $p_J^0=|\bold{p_J}|=|\bold{p}|$. Thus, the partonic momentum conservation can be written as
\bea
p_I^\mu = q^\mu + k_{out}^\mu + \sum_i^N p_{Ji}^{\mu}  + v^\mu \big [ \sum_i^N (p^{c+}_i+k_i^+)\big ],
\eea
which is equivalent to the condition
\bea
1-z=\frac{2}{\hat{s}} \Big [ p_I \cdot k_{out} + p_I^0 \sum_{i}^N (p_i^{c+} +k_i^+)\Big ] + {\cal O}(\lambda^4).
\eea
Thus, the limit of partonic threshold constrains the energy component of soft radiation outside the jets ($k_{out}$) and the null components of the total jet momenta ($p_i^{c+} +k_i^+$).  We note that the threshold variable $z$ must be defined carefully in 
order to avoid the appearance of non-global logarithms~\cite{Banfi:2008qs}; the definition above is expected to not contain such terms~\cite{Bauer:2010vu}.
 
We note that there can be configurations consistent with Eq.~(\ref{meff}) where one or more of the N-jets becomes soft, so that there is a hierarchy between the different transverse momenta $p_{J}^T$. However, the inclusive nature of the observable we consider forces such configurations to occupy a small corner of phase space.  They are consequently phase-space suppressed. This is the same phase-space suppression that makes fully inclusive Drell-Yan processes insensitive to special exclusive jet configurations. In our case, we are interested in the tail of the $m_{eff}$ distribution so that $m_{eff} > 1$ TeV.  Following the experimental studies, we further restrict $p^T_J >p^T_{min}$ with $p^T_{min} \approx 100$ GeV, and impose standard cuts demanding that the photon and jets are well separated.  Two types of regions can be identified. In the first region, all the jet momenta are of the same order so that $|{p}^T_{J_i}| \sim |{p}^T_{J_j}|$. A second type of region can arise where one or more of the jets becomes soft so that hierarchies such as $|{p}^T_{J_i}| \gg |{p}^T_{J_j}|$ can arise.  This second type of region is suppressed for several reasons.  First, 
since $p^T_{min} \ll m_{eff}$, events with widely disparate jet momenta populate only a small corner of the full phase space.  This can be understood by noting that if one of the jets is soft, then the integration measure over the corresponding jet transverse momentum scales like $(p^T_{min}/m_{eff})^2$. Similarly, the integration measure over its rapidity is also suppressed in order to maintain the requirement that the jet is soft.  The allowed phase space of the remaining hard jets is also further restricted to the high transverse-momentum region by the requirement that the event have large $m_{eff}$, which must now be accomplished with a fewer number of states.  In other words, since we are inclusively integrating over each jet transverse momentum up to the maximum allowed valued determined by $m_{eff}$, we are not sensitive to special configurations that occupy a small corner of phase space. Another source of suppression comes from the hard production amplitude. One or more low-$p_T$ jets require that the remaining jets and gauge boson have very high $p_T$ in order to produce a large $m_{eff}$.  This 
configuration is suppressed by the steep fall-off of the matrix elements at high transverse momentum of the leading jets.  

\begin{figure}[h!]
\begin{minipage}[b]{3.0in}
  \includegraphics[width=2.8in,angle=90]{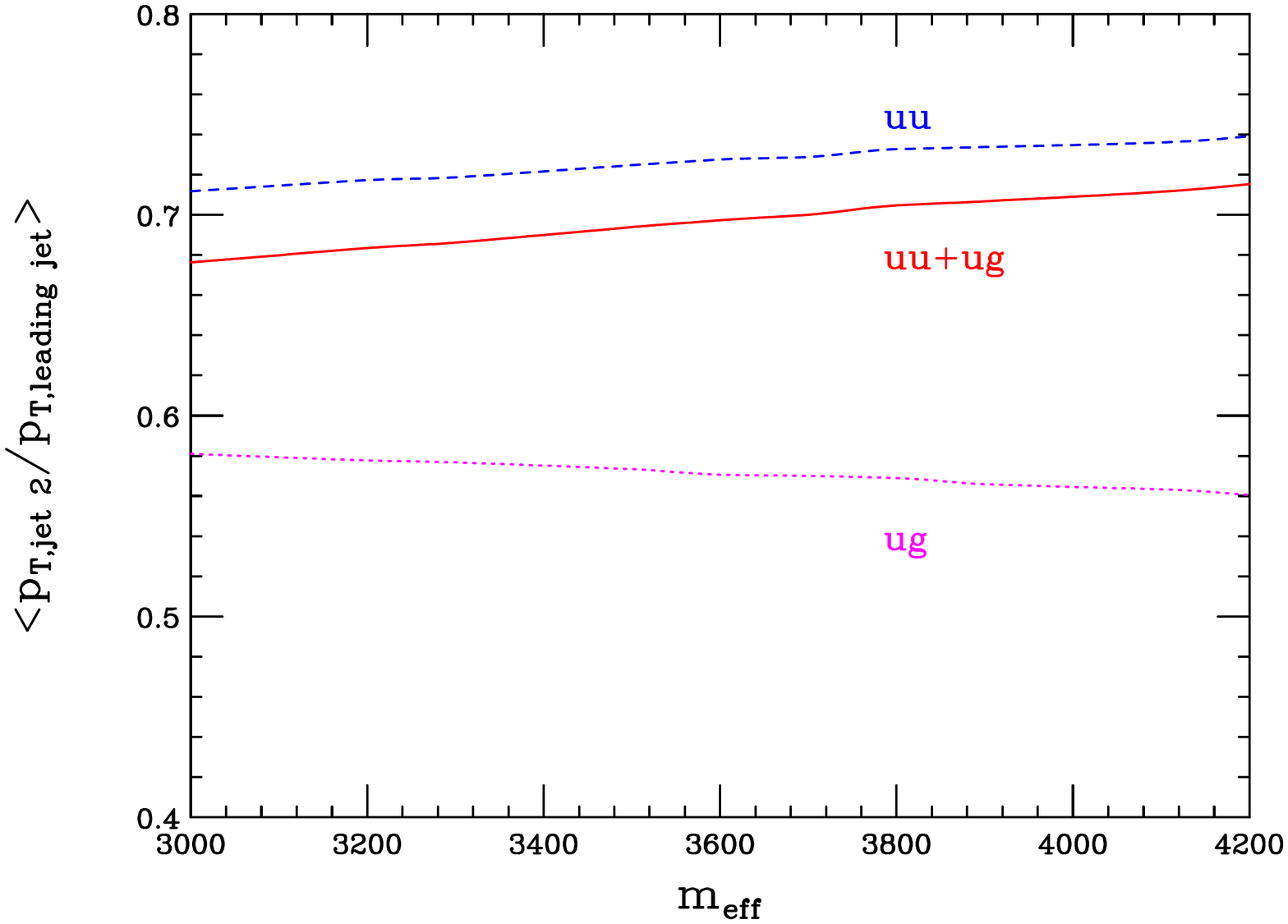}
\end{minipage}
\begin{minipage}[b]{3.0in}
  \includegraphics[width=2.8in,angle=90]{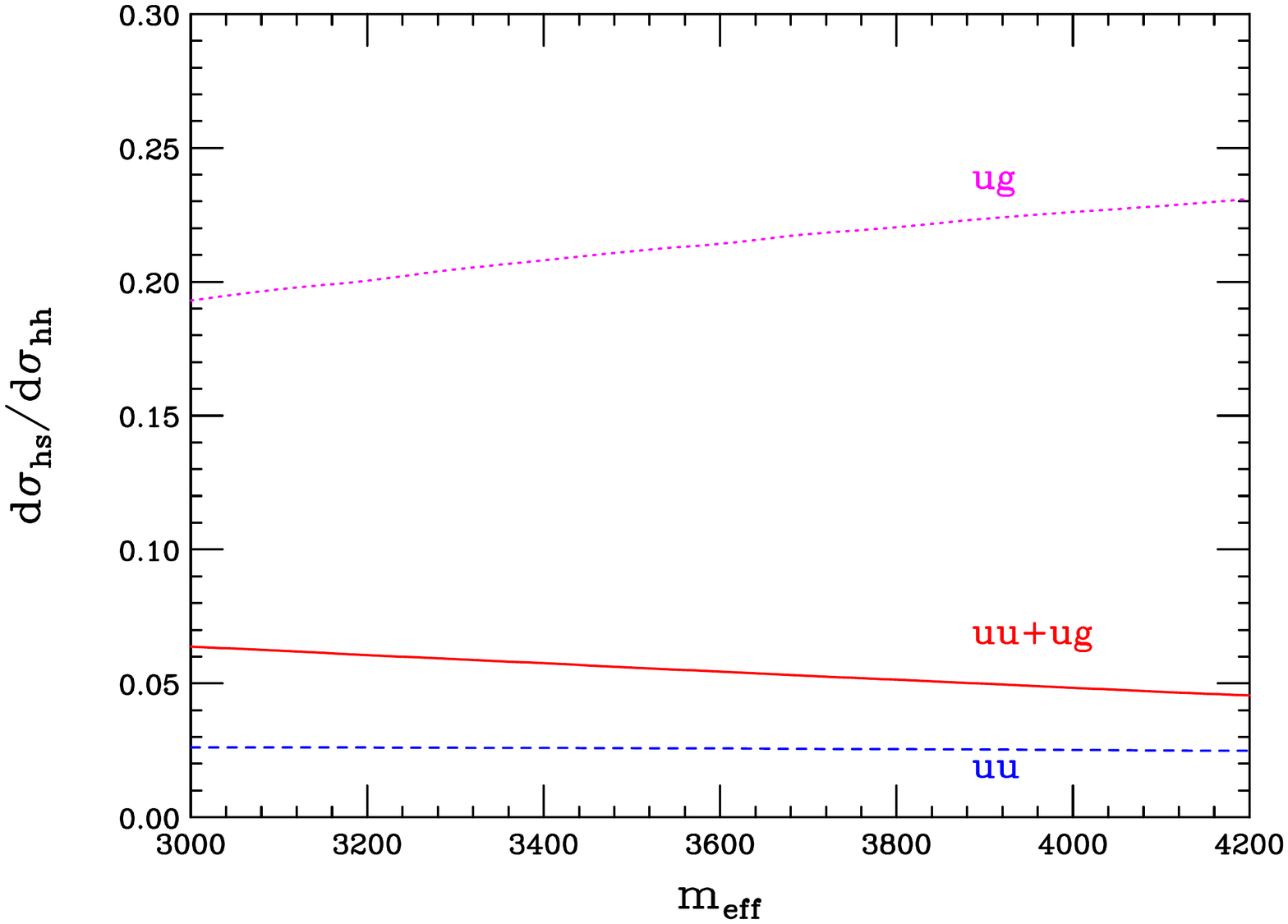}
\end{minipage}
\vspace{-0.5cm}
\caption{The left panel shows the average ratio of the second jet $p_T$ over the $p_T$ of the leading jet for the $ug$ partonic channel (magenta dotted line), the $uu$ partonic channel (blue dashed line), and for the total result (red solid line).  The right panel shows the ratio of the cross section for one hard jet and one soft jet over the two hard-jet cross section, where the soft jet is defined by $100 \,\text{GeV} \leq p^T_2 \leq 0.2 \times m_{eff}/2$.  The double-hard region is then defined as the difference of the cross section with 
both transverse momenta greater than 100 GeV, minus the cross section with either jet having a $p_T$ between 100 GeV and $0.2 \times m_{eff}/2$.  These results use the LO approximation of the cross section.}
\label{jetphasespace}
\end{figure}

For illustration we show numerically the suppression of these configurations for the process $pp \to \gamma + 2$ jets in Fig.~\ref{jetphasespace}.  In the left panel the average ratio of the sub-leading jet $p_T$ over the leading jet $p_T$ is plotted.  This ratio is approximately 0.7 in the high $m_{eff}$ region, indicating that the jets have roughly the same $p_T$ on average.   Further evidence is shown in the right panel, where a soft-jet region of phase space is defined by requiring $100 \, \text{GeV} \leq p^T_2 \leq 0.2 \times m_{eff}/2$.  The contribution of one soft jet and one hard jet to the high-$m_{eff}$ distribution is 
about 5\% of the contribution from two hard jets, indicating that it contributes only a small amount to the cross section.  These numerical results are in fact a conservative estimate of the suppression of the one soft-jet region, since a portion of the region with a low-$p_T$ second jet includes a second jet that is energetic but at higher rapidity.  The region of large $m_{eff}$ is therefore dominated by the production a gauge boson and N-hard jets near threshold.  The dynamics of this configuration is best described by N-collinear sectors and soft radiation with threshold kinematics as described in this section, and is well-suited to the application of the SCET 
formalism used here.

\section{SCET operator basis}
\label{sec:scetop}

The process $pp\to \gamma + 2 $ jets receives contributions from
various partonic channels, which we enumerate here.  The channels with four participating quarks or antiquarks are given by 
\bea
\label{qqqq}
q q' \to qq'\gamma, \qquad \bar{q}\bar{q}' \to  \bar{q}\bar{q}' \gamma, \qquad q\bar{q}' \to q\bar{q}' \gamma, 
\eea
and the channels with two quarks or antiquarks and two gluons are given by
\bea
\label{qqgg}
qg\to qg\gamma, \qquad \bar{q} g \to \bar{q} g \gamma ,\qquad q\bar{q} \to gg \gamma, \qquad gg \to q\bar{q} \gamma. 
\eea
Contributions where the two jets are produced by the decay of an electroweak gauge boson are expected to be small, and are neglected.  The processes above are mediated by operators in SCET with Wilson coefficients determined by a matching calculation from QCD to SCET. The SCET operators can be decomposed in terms of their color structures. There are two color structures for the four-quark operators,
\bea
\label{ops1}
\theta_1^{\alpha \beta \gamma \delta} &=& (\bar{\chi}_2^\alpha\> t^a \>\chi_1^\beta) \>( \bar{\chi}_4^\gamma \>t^a \>\chi_3^\delta), \nn \\
\theta_2^{\alpha \beta \gamma \delta} &=& (\bar{\chi}_2^\alpha\> \textbf{1} \>\chi_1^\beta) \>( \bar{\chi}_4^\gamma \>\textbf{1} \>\chi_3^\delta),
\eea
and three color structures for the operators with two quark fields and two gluon fields:
\bea
\label{ops2}
\Theta_{1,\mu \nu}^{\alpha \beta} &=& (\bar{\chi}_2^\alpha\> t^{a_1} t^{a_3} \>\chi_4^\beta)\> A^{a_1}_\mu A^{a_3}_\nu, \nn \\
\Theta_{2,\mu \nu}^{\alpha \beta} &=& (\bar{\chi}_2^\alpha \>t^{a_3} t^{a_1} \>\chi_4^\beta) \>A^{a_1}_\mu A^{a_3}_\nu, \nn \\
\Theta_{3,\mu \nu}^{\alpha \beta} &=& (\bar{\chi}_2^\alpha \>\delta^{a_1 a_3} \>\chi_4^\beta)\> A^{a_1}_\mu A^{a_3}_\nu.
\eea
The Dirac indices are denoted by $\{ \alpha, \beta , \gamma, \delta \}$, with the remaining Greek indices denoting Lorentz indices. The amplitudes for the four-quark processes take the form
\bea
\label{Mtheta}
{\cal M}_\theta = \sum_{I=1}^2 \epsilon^*_\mu \>C_{\alpha \beta \gamma \delta}^{I,\mu} \>\langle \theta_I^{\alpha \beta \gamma \delta} \rangle ,
\eea
where $\epsilon_\mu$ is the photon polarization vector, $C_{\alpha \beta \gamma \delta}^{I,\mu}$ is the matching coefficient that also includes the spin structure, and $\langle \theta_I^{\alpha \beta \gamma \delta} \rangle$ denotes the matrix element of the corresponding  SCET operator. Similarly, the amplitude for the channels with two gluons is given by
\bea
\label{MTheta}
{\cal M}_\Theta = \sum_{I=1}^3 \epsilon^*_\rho \>C_{\alpha \beta }^{I,\rho \mu \nu} \>\langle \Theta_{I,\mu \nu}^{\alpha \beta } \rangle.
\eea
The Wilson coefficients in Eqs.~(\ref{Mtheta}) and~(\ref{MTheta}) depend on the specific partonic channel in Eqs.~(\ref{qqqq}) and~(\ref{qqgg}) under consideration.

\section{Factorization}
\label{sec:fac}

In the threshold limit, the process $pp\to \gamma + $N jets is characterized by N-narrow jets and only soft radiation outside of these jets. The dynamics of such a process can be described in terms of collinear degrees of freedom along the jet directions, soft degrees of freedom throughout the event, and the initial-state PDFs. The factorization formula takes the schematic form
\bea
\label{schem}
d\sigma \sim H_{IJ} \otimes S_{JI}\otimes J_1\otimes \cdots \otimes J_N  \otimes f\otimes f,
\eea
where $\otimes$ denotes a convolution structure, $H_{IJ}$ denotes the hard function and  $S_{JI}$ denotes the soft function with the indices $I,J$ running over the color structure basis in Eqs.~(\ref{ops1}) and~(\ref{ops2}). The factors of $J_i$ correspond to the collinear jet functions and $f$ denotes the PDF. The hard function encodes the physics of the hard partonic interaction, the jet functions describe the dynamics of the collinear momenta in the jet regions, and the soft function describes the soft radiation in the event which is either grouped in one of the jet regions or outside of the jet regions as in Eq.~(\ref{softmom}). The momentum scalings of the soft and collinear degrees of freedom in the threshold limit are given in Eq.~(\ref{scaling}).  The relevant scales in the problem can be characterized by
\bea
\sqrt{\hat{s}} \gg \sqrt{\hat{s}} \sqrt{1-z} \gg \sqrt{\hat{s}} (1-z) \gg \Lambda_{QCD},
\eea
so that the hard function, the jet functions, and the soft function are evaluated at the typical scales $\mu_H \sim \sqrt{\hat{s}}$, $\mu_J \sim \sqrt{\hat{s}} \sqrt{1-z} $,  and $\mu_S \sim \sqrt{\hat{s}} (1-z)$, respectively. The optimal choice of scales for the most stable resummation can be affected by the shape of the luminosity function ~\cite{Appell:1988ie,Becher:2007ty}.  This is discussed in the next section.

\begin{figure}[h!]
\begin{center}
\includegraphics[width=3.5in]{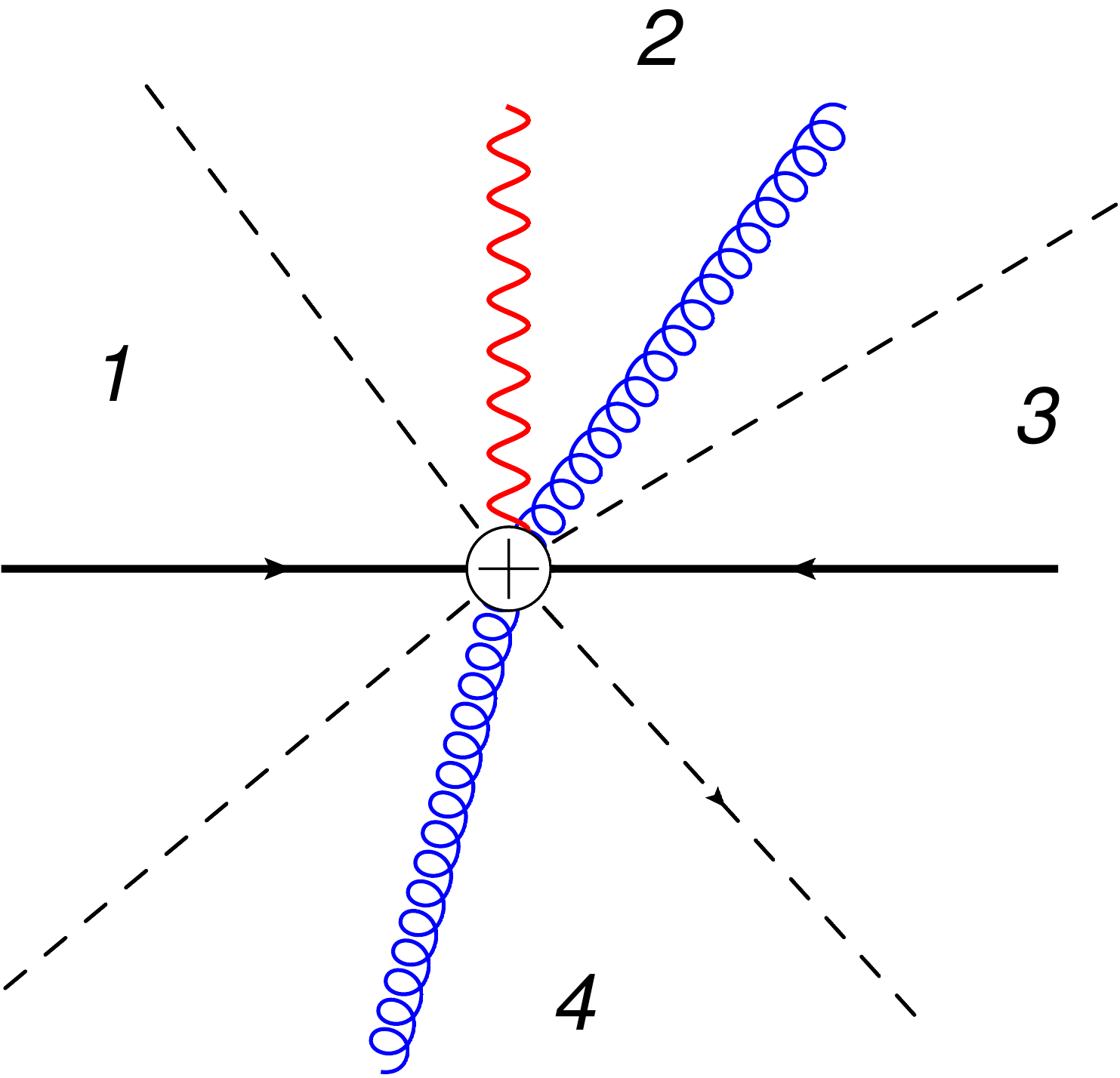}
\end{center}
\caption{The four jettiness regions into which each event is divided, illustrated using the $q\bar{q} \to gg \gamma$ process.  The dashed lines indicate the separation between the regions.  Regions 1 and 3 correspond to the beam directions, while regions 2 and 4 correspond to the two jets.}
\label{regions}
\end{figure}

In the N-jettiness formalism, another scale that appears is related to $\tau_N$ or equivalently the $\tau_N^i$ in Eq.~(\ref{tauN}). The $\tau_N^i$ correspond to the contribution to the total N-jettiness from region-$i$ in the event.
For the process $pp\to \gamma +$2 jets, this is divided into four distinct regions as shown in Fig.~\ref{regions}. Regions 1 and 3 correspond to the two beam directions, while regions 2 and 4 correspond to the two jet regions. We give a factorization formula that is differential in the  quantity
\bea
\hat{\tau}_2 \equiv \tau_2^{J_2} + \tau_2^{J_4}.
\eea
where $\tau_2^{J_2}$ and $\tau_2^{J_4}$ correspond to the contribution to the 2-jettiness from the regions of jet-2 and jet-4 respectively. Note that $\hat{\tau}_2$ is distinct from $\tau_2$ since it does not include the contributions from the soft radiation in regions 1 and 3 which lie outside the two jet regions. In the threshold limit, $\hat{\tau}_2\sim \sqrt{\hat{s}} \lambda^2$, corresponding to the size of the jet invariant masses. Thus, $\hat{\tau}_2$ has a size that corresponds to the typical soft or residual momenta in the event. 

The factorization formula takes the general form
\bea
\label{fac}
\frac{\mathrm{d}\sigma}{\mathrm{d}\Phi_{q_{J_2}}\, \mathrm{d}\Phi_{q_{J_4}}\,  \mathrm{d}\Phi_q \mathrm{d}\hat{\tau}_2 \,} & = & \,
 \frac{(2\pi)^4}{Q^4\tau}\,
 \int_\tau^1 \mathrm{d}z \,
\int_{-\frac{1}{2}\ln\frac{z}{\tau}}^{\frac{1}{2}\ln\frac{z}{\tau}} \mathrm{d} y \,
 \int \mathrm{d}s_{J_2} \,
\mathrm{d}s_{J_4} \int \mathrm{d} k_{out}^0 \int  \mathrm{d}\tau_{2,s}^{J_2}\,\mathrm{d}\tau_{2,s}^{J_4}\,
\nn \\
&& \times \delta^2(q_{J_2}^T +  q_{J_4}^T +q^T) \,
\delta\left(y - \tanh^{-1}\left(\frac{q_{J_2}^z+q_{J_4}^z+q^z}{q_{J_2}^0+q_{J_4}^0+q^0}\right)\right)\, \nn \\
&&\times  \delta\left(1-z - \frac{2\cosh y}{\sqrt{\hat{s}}}  \,
\hat{\tau}_2 - \frac{2}{\sqrt{\hat{s}}}k_{out}^0 \right)\,
 \delta\left(\hat{\tau}_2 - \frac{s_{J_2}}{\bar{q}_{J_2}} - \frac{s_{J_4}}{\bar{q}_{J_4}}\,
 - \tau_{2,s}^{J_2} - \tau_{2,s}^{J_4}\right) \, \nn \\
&&\times H_{IJ}(\bar{q}_i,n_i,\mu_S;\mu_H)\, S_{JI}(k_{out}^0,\tau^{J_2}_{2,s},\tau^{J_4}_{2,s},\mu_S) \nn \\
&&\times f(x_a,\mu_S)\, f(x_b,\mu_S)\, J_{J_2}(s_{J_2},\mu_S;\mu_J) \, J_{J_4}(s_{J_4},\mu_S;\mu_J) \,,
\eea
where we have defined $k_{out}^0 = k_{out}\cdot p_I/|p_I|$.
The cross-section above is fully differential in the massless jet momenta $q_{J_2}$ and $q_{J_4}$ defined in Eq.~(\ref{pj}), the photon momentum $q$, and $\hat{\tau}_2$.  We note that $d\Phi_i$ denotes the phase space of the massless particle $i$,
\begin{equation}
d\Phi_i = \frac{d^3 p_i}{2 (2\pi)^3 E_i}.
\end{equation}
The initial partonic momentum fractions $x_{a,b}$ are given by
\bea
\label{xfrac}
x_a = \sqrt{\frac{\tau}{z}}e^y\,, \hspace{5.ex} x_b = \sqrt{\frac{\tau}{z}}e^{-y}.
\eea
The arguments $\bar{q}_i$ of the hard function denote the projection of the total final-state momentum onto the direction $n_i$, with $i$ denoting any of the four regions defined above.  This factorization formula can be easily generalized for processes with more than two jets. As shown in the  schematic form given in Eq.~(\ref{schem}), for the case of N-jets the above formula will be modified to have  an appropriately generalized soft function convoluted with N jet functions. The $m_{eff}$ distribution can be obtained from the factorization formula above by performing the integrations over the remaining phase space factors and over $\hat{\tau}_2$ after inserting the delta function constraint $\delta(m_{eff} -q_{J_2}^T -  q_{J_4}^T -q^T )$.

The specific form of the hard, jet, and soft functions in Eq.~(\ref{fac}) will depend on the partonic channel that mediates the process $pp\to \gamma +2$ jets. For example, for the partonic channels where the final-state jets are initiated by quarks, the factorization formula will be in terms of quark jet functions. Similarly, processes with jets initiated by gluons will involve gluon jet functions. The color representations of the Wilson lines in the soft function will also depend on the partonic channel. Similarly, the PDFs in Eq.~(\ref{fac}) will correspond to that of the initial-state partons in the partonic process. Finally, the hard function must be computed separately for each partonic channel, and is given by the spin-summed and color-ordered partial amplitude squared. 

The field theoretic definitions of the different types jet and soft functions are listed in the Appendix for completeness. For the partonic channels in Eq.~(\ref{qqqq}) and Eq.~(\ref{qqgg}), mediated by the operators defined in Eqs.~(\ref{ops1}) and~(\ref{ops2}) respectively, the hard function takes the form 
\bea
\label{hard1}
H_{IJ}^\theta =  -g_{\mu \mu'}C_{\alpha \beta \gamma \delta}^{I,\mu}C_{\alpha ' \beta '\gamma ' \delta '}^{*J,\mu '} (\qslash_2)^{\alpha \alpha '} (\qslash_4)^{\gamma \gamma '} (\qslash_1)^{\beta \beta '} (\qslash_3)^{\delta \delta '}.
\eea
and 
\bea
\label{hard2}
H_{IJ}^\Theta = -g_{\rho \rho '}g_{\mu \mu '}g_{\nu \nu '} C_{\alpha \beta }^{I,\rho \mu \nu}C_{\alpha ' \beta '}^{*J, \rho '\mu ' \nu '} (\qslash_2)^{\alpha \alpha '} (\qslash_4)^{\beta \beta '}.
\eea
The Wilson coefficients $C_{\alpha \beta \gamma \delta}^{I,\mu}$ and $C_{\alpha \beta }^{I,\rho \mu \nu}$ are defined in Eqs.~(\ref{Mtheta}) and~(\ref{MTheta}) and depend on the partonic channel being considered. The momenta $q_i$ in Eqs.~(\ref{hard1}) and~(\ref{hard2}) correspond to the label momenta of the collinear fields in Eqs.~(\ref{ops1}) and~(\ref{ops2}). The tree-level expressions for these hard functions for the various partonic channels are given in the Appendix. We note that the hard functions are known at the one-loop level~\cite{Kunszt:1994tq,Bern:1994fz}.

In Eq.~(\ref{fac}), large logarithms are summed via renormalization group (RG) equations for the hard, jet, and soft functions, while the PDFs are evolved via the standard DGLAP equations. In Eq.~(\ref{fac}) large logarithms are summed by running hard function from the hard scale $\mu_H$ to the soft scale $\mu_S$, while the jet function is run from the jet scale $\mu_J$ to the soft scale $\mu_S$.  The soft function and the PDFs are evaluated at the soft scale $\mu_S$. The RG evolution equations for the hard and jet functions  are given in the Appendix for completeness.

\section{Numerical Results}
\label{sec:num}

\subsection{Explicit form of the factorization formula at NLL}

\begin{figure}[h!]
\begin{center}
\includegraphics[width=4.0in,angle=90]{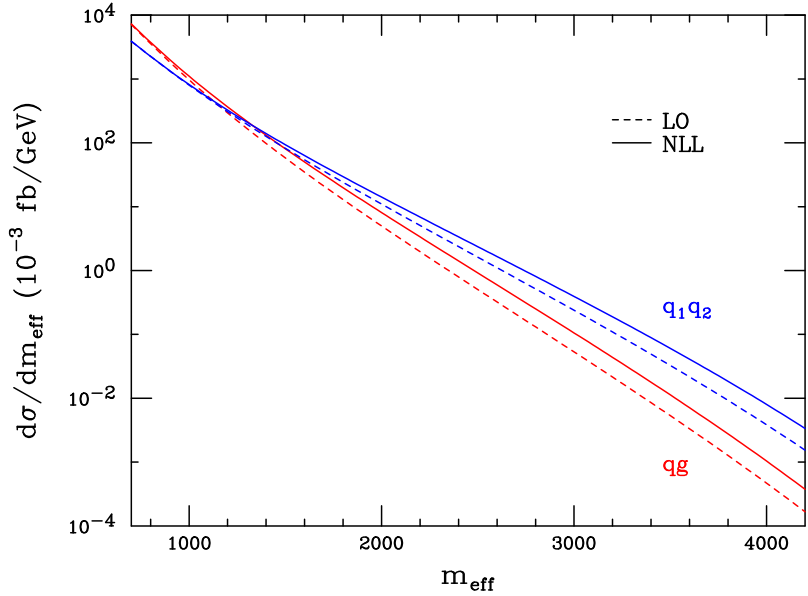}
\end{center}
\vspace{-0.7cm}
\caption{The solid and dashed lines correspond to the $m_{eff}$ distributions with NLL resummation and at LO, respectively.  The $qg$ initial state is shown in red while the $q_1 q_2$ initial state, with $q_i$ representing any quark or anti-quark, is shown in blue.}
\label{partonicmeff}
\end{figure}

In this section we give numerical results for the $m_{eff}$ distribution for the $pp\to \gamma+2$ jets process. In particular, we present results at the NLL level of accuracy. We follow the conventions outlined in Table 1 of Ref.~\cite{Berger:2010xi} for the counting of logarithms. At the NLL level of accuracy, the factorization formula in Eq.~(\ref{fac}) can be computed using tree-level values for the hard, jet, and soft functions. The RG evolution is performed by running the hard function between the hard and soft scales $\mu_H$ and $\mu_S$. The jet function is run between the jet and soft scales $\mu_J$ and $\mu_S$. The soft function and the PDFs are evaluated at the soft scale.  The RG evolution of the hard and jet functions at NLL accuracy requires the corresponding two-loop and one-loop  cusp and non-cusp anomalous dimensions respectively. Finally, at NLL the LO PDFs are DGLAP-evolved using two-loop running of the strong coupling.  After integrating Eq.~(\ref{fac}) over $\hat{\tau}_2$ in the range $[0,\hat{\tau}^{\rm cut}_2]$, the explicit form of the cross section at the NLL level of accuracy is given by 
\bea
\label{facNLL}
\frac{\mathrm{d}\sigma^{\text{NLL}} (pp\to \gamma+2\text{ jets})}{\mathrm{d}\Phi_{q_{J_2}}\, \mathrm{d}\Phi_{q_{J_4}}\,  \mathrm{d}\Phi_q}  &=&
\frac{(2\pi)^4}{Q^4\tau}\int_{z_{min}}^1 \mathrm{d}z \,
 \delta^2(q_{J_2}^T +  q_{J_4}^T +q^T)\,
\text{Tr}\left[\tilde{H}^{\text{NLL}}(\mu_S;\mu_H)\tilde{S}^{(0)}\right] \nn \\
&&\times \exp\left(-4(C_{J_2}+C_{J_4})S(\mu_J,\mu_S)\,
-A_{J_2}(\mu_J,\mu_S)-A_{J_4}(\mu_J,\mu_S)\right)\nn\\
&&\times \left(\frac{\mu_H}{\bar{q}_{J_2}}\right)^{-2C_{J_2}A_\gamma(\mu_J,\mu_S)}\,
 \left(\frac{\mu_H}{\bar{q}_{J_4}}\right)^{-2C_{J_4}A_\gamma(\mu_J,\mu_S)}\nn \\
&&\times \,
\left(\frac{2 \mu_J^2 \cosh y }{\mu_H\sqrt{\hat{s}}}\right)^{\w}
\frac{\left(e^{\gamma_E}\right)^{\w}}{\Gamma(-\w)}\,
\left(\frac{1}{1-z}\right)^{1+\w}  f_a(x_a)f_b(x_b)\,,
\eea
%
where the dependence of $\hat{\tau}_2^{\rm cut}$ is implicit in the definition of $z_{min}$.  We have used the result of Eq.~(\ref{jmult}) to include the RG evolution of the jet functions in the above expression.  We now explain the various parts of the above formula. The matrix $\tilde{H}^{\text{NLL}}(\mu_S;\mu_H)$ denotes the tree-level hard function in a color-rotated basis with NLL RG evolution between $\mu_H$ and $\mu_S$. In this color-rotated basis, the evolution of the hard function matrix elements is multiplicative, as seen in Eq.~(\ref{Hrun}). The relation between the original hard-function matrix $H$ obtained from matching QCD onto SCET operators is related to the function $\tilde{H}$ in the rotated basis as shown in Eq.~(\ref{Hrot}). Similarly, $\tilde{S}^{(0)}$ denotes the tree-level soft function in the same color-rotated basis. The partonic momentum fractions $x_{a,b}$ for the initial-state partons of flavor $a,b$ are given by Eq.~(\ref{xfrac}). The parameter $\w$ is defined as
\bea
\label{w}
\w = -2(C_{J_2}+C_{J_4})A_\gamma(\mu_J,\mu_S),
\eea
and all the remaining quantities in Eq.~(\ref{facNLL}) and on the right-hand side of Eq.~(\ref{w}) are defined in the Appendix. Note that for $\mu_H> \mu_S$, $\omega < 0$, so that the plus-prescription for the $z\to 1$ limit has been dropped in Eq.~(\ref{facNLL}). The rapidity $y$ of the entire two-jet and photon system in Eq.~(\ref{facNLL}) is given by
\bea
y=\tanh^{-1}\frac{q_{J_2}^z+q_{J_4}^z+q^z}{q_{J_2}^0 + q_{J_4}^0 + q^0}.
\eea
$z_{min}$ is determined by
\bea
\label{zmin}
z_{min} = \text{max}\Big [1-\frac{1}{2}A(\sqrt{A^2+4}-A),\tau\exp(2|y|)\Big ]\,,
\eea
where 
\bea
A =  \frac{2\cosh y \>\hat{\tau}_2^{\rm cut}}{\sqrt{\tau s}}.
\eea
We choose $\hat{\tau}_2^{\rm cut}$ to be of the order of the soft scale so that $\hat{\tau}_2^{\rm cut} \sim \mu_S \sim \sqrt{\hat{s}} (1-z)$.  For a fixed $\hat{\tau}_2^{\rm cut}$, in the machine-threshold limit $\tau \to 1$, $z_{min}$ in given by the second argument in Eq.~(\ref{zmin}) and is independent of $\hat{\tau}_2^{\rm cut}$. This is simply understood by the fact that in the threshold limit, the jets become extremely narrow and thus insensitive to $\hat{\tau}_2^{\rm cut}$. In other words, for a fixed $\hat{\tau}^{\rm cut}_2$ the threshold condition eventually becomes a stronger constraint on the jet masses as compared to  $\hat{\tau}_2^{\rm cut}$ when $\tau \to 1$. For the numerical results, we  choose $\hat{\tau}_2^{\rm cut}$ to be large enough so that $z_{min}$ is independent of it.  If we were 
to choose a small $\hat{\tau}_2^{\rm cut}$, logarithms  of this quantity associated with vetoing additional jets would appear in the fixed-order result.  Since $\hat{\tau}_2$ is related to $1-z$ through the delta-function constraint in Eq.~(\ref{fac}), our factorization formula also 
provides theoretical control over logarithms of this variable.  This is seen explicitly in the NLL result above by the relation between $\hat{\tau}_2^{\rm cut}$ and $z_{min}$.  We integrate over the phase space factors in Eq.~(\ref{facNLL}) with the delta-function constraint $\delta[m_{eff} - q_{J_2}^T - q_{J_4}^T - q^T]$ in order to generate numerical results for the $m_{eff}$ distribution.

\begin{figure}[h!]
\begin{center}
\includegraphics[width=4in,angle=90]{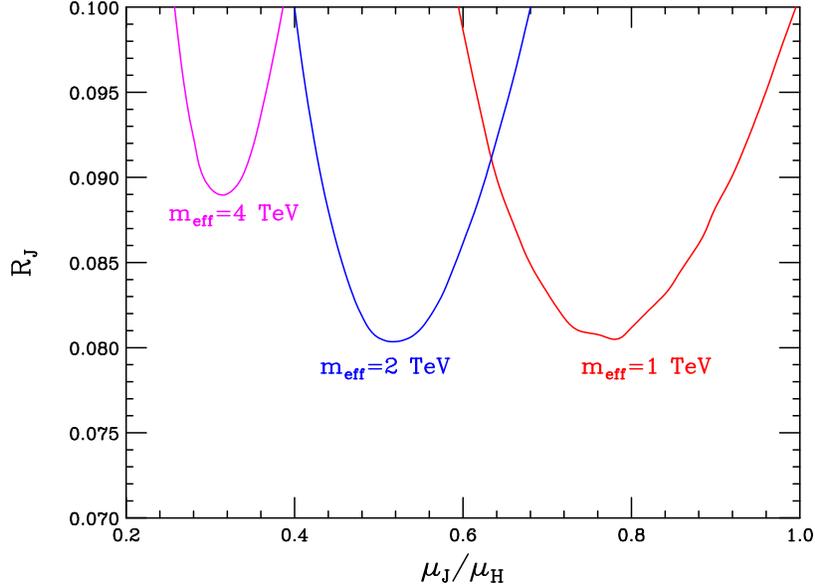}
\vspace{-0.7cm}
\caption{The dependence of the ratio $R_J$ on the choice of the jet scale $\mu_J$ for various kinematic points. From the right to the left, are the
curves used to determine the jet scales for 
$m_{eff}$ = 1 TeV, 2 TeV and 4 TeV, respectively.}
\label{scaledetermine}
\end{center}
\end{figure}

\subsection{Numerics for a 7 TeV LHC}

We begin by showing the $m_{eff}$ distributions for the important partonic channels in Fig.~\ref{partonicmeff}. The dominant contributions come from the $q_1 q_2$ and $qg$ initial states, where the $q_i$ represent any quark or anti-quark.  We note that the $gg$ initial state contributes only at the percent level or less for the considered $m_{eff}$ range.  The dotted and solid curves correspond to the LO and NLL results respectively for the different partonic channels as described in the caption.  Lower cuts of 100 GeV have been imposed on both jets and on the photon, and all final-state particles are required to be separated by $\Delta R_{ij} >0.4$ 
with $\Delta R_{ij}^2 = (\eta_i - \eta_j)^2+(\phi_1-\phi_2)^2$.  CTEQ6 LO PDFs~\cite{Pumplin:2002vw} have been used to produce these numerical results.  In order to obtain the LO curves we choose the scales 
\begin{equation}
\mu_H=\mu_J=\mu_S =\sqrt{q_{J_2}^T q_{J_4}^T},
\end{equation}
effectively removing the resummation.  This dynamical scale is similar to scales found to reduce the effect of higher-order corrections in fixed-order calculations of $W$+multi-jet production~\cite{Berger:2009zg}.  For the NLL curves, we choose the hard scale as $\mu_H=\sqrt{q_{J_2}^T q_{J_4}^T}$. The soft scale is set by the see-saw relation $\mu_S = \mu_H^2/\mu_J$. The jet scale $\mu_J$ is determined numerically by minimizing the contribution of the logarithmic terms in the NLO jet function to the cross-section. In particular, we minimiz the quantity
\bea
\label{jetratio}
R_J = \frac{d\sigma^{NLO_{jet-logs}}(\mu_H=\mu_S=\mu_J)}{d\sigma^{LO}(\mu_H=\mu_J=\mu_S =\sqrt{q_{J_2}^T q_{J_4}^T})}
\eea
with respect to $\mu_J$. The numerator is defined as the contribution to the cross-section only from the logarithmic terms in the NLO jet functions. In Fig.~\ref{scaledetermine}, we show the dependence of the ratio $R_J$ on $\mu_J$ for several representative values of $m_{eff}$. We see that clearly identifiable minima occur for specific choices of $\mu_J$. As $m_{eff}$ increases the ratio $\mu_J/\mu_H$ becomes smaller, indicating a growing hierarchy between the hard and jet scales. As $m_{eff}$ decreases, we see that the ratio $\mu_J/\mu_H$ becomes comparable to one, indicating that there is no longer a large hierarchy. In this case the effect of threshold resummation becomes smaller, and one can rely on fixed-order perturbation theory. For the numerical results presented in this section, the jet scale is determined numerically for each value of $m_{eff}$. We have tested that choosing other values of the hard scale, such as $\mu_H = m_{eff}/3$, leads to negligible numerical differences.

\begin{figure}[h!]
\begin{center}
\includegraphics[width=4.0in,angle=90]{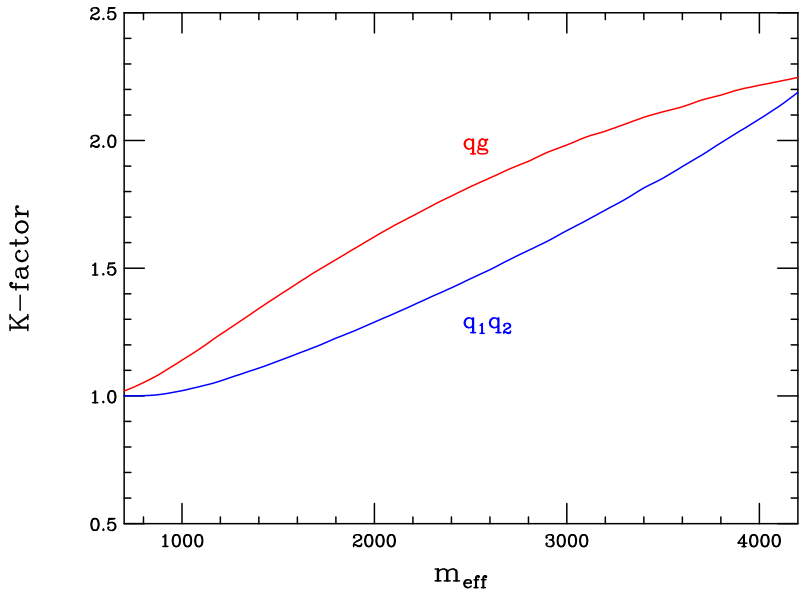}
\vspace{-0.7cm}
\caption{A plot of the K-factor, defined as the ratio of the cross section with NLL resummation over  the LO result, for the important partonic channels. The curves from top to bottom correspond to the $qg$ and $q_1 q_2$ initial states, where $q_i$ denotes any quark or anti-quark.}
\label{Kfactor}
\end{center}
\end{figure}

\begin{figure}
\begin{center}
\includegraphics[width=4in,angle=90]{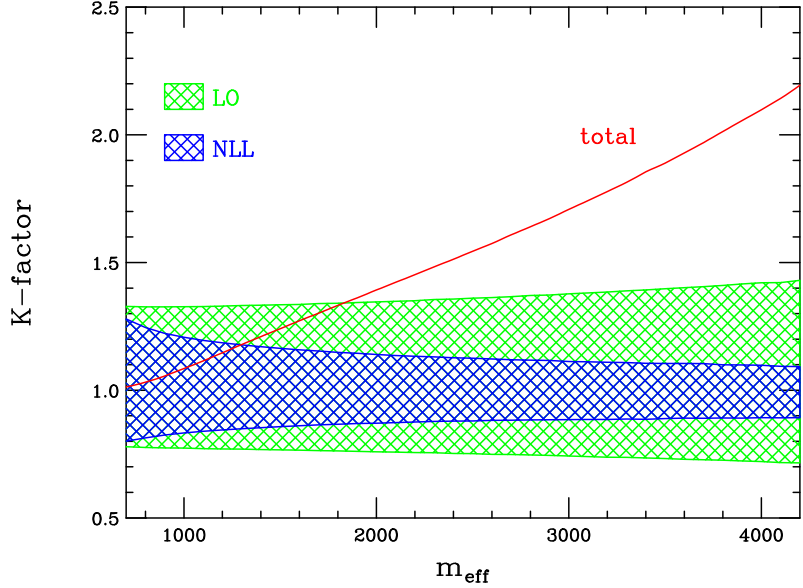}
\vspace{-0.7cm}
\caption{Plot showing the total K-factor defined as the ratio of the NLL and LO cross-section as a function of $m_{eff}$.  Also shown are the scale dependences of the LO and NLL results.  The range of variation is defined in the text.}
\end{center}
\label{Kfactor-scale}
\end{figure}

From Fig.~\ref{partonicmeff} we see that the effect of resummation becomes more important in the region of large $m_{eff}$, which is dominated by threshold kinematics. This is further illustrated in Fig.~\ref{Kfactor}, where we show the K-factor for the same partonic channels as a function of $m_{eff}$. The K-factor is defined as the ratio of the cross-section with NLL resummation over the LO cross-section. 
Finally, in Fig. ~\ref{Kfactor-scale} we show the total K-factor and the uncertainties associated with scale variation. The solid red curve shows the total K-factor as a function of $m_{eff}$. The wider green band is obtained by setting $\mu_H=\mu_J=\mu_S=\mu$ and varying the scale $\mu$ in the range $\{ 1/2,2\}$ around the central value $\mu=\sqrt{q_{J_2}^T q_{J_4}^T}$ for the ratio
\bea
K_{LO} = \frac{d\sigma^{LO} (\mu)}{d\sigma^{LO} (\mu^c=\sqrt{q_{J_2}^T q_{J_4}^T})}.
\eea
The narrower blue band is the result of scale variation of the NLL resumed cross-section. In this case, we vary the scales $\mu_H,\mu_J,\mu_S$ in the range $\{1/2,2\}$ around their central values.  The central value of the hard scale is taken as $\mu_H= \sqrt{q_{J_2}^T q_{J_4}^T}$.  The determination of the jet and soft-scale central values was described earlier.  More explicitly, the NLL scale variation band is determined by varying the scales $\mu_H,\mu_J,\mu_S$ in the ratio
\bea
K_{NLL} = \frac{d\sigma^{NLL} (\mu_H,\mu_J,\mu_S)}{d\sigma^{NLL} (\mu_H^c=\sqrt{q_{J_2}^T q_{J_4}^T},\mu_J^c,\mu_S^c )},
\eea
where again we have normalized with the NLL cross-section evaluated at the central scale choices $\mu_H^c,\mu_J^c,\mu_S^c$. We see from Fig.~\ref{Kfactor-scale} that scale variation uncertainty is significantly reduced when NLL resummation is included.

\section{Conclusions}
\label{sec:conc}

In this manuscript we have begun an investigation of the effect of threshold logarithms in gauge boson plus multi-jet production at the LHC.  Such processes serve as backgrounds to supersymmetric particle production and to other forms of new physics that contain 
missing energy signatures.  We have focused on $pp \to \gamma + 2$ jet production, which is used as a calibration process for 
the production of missing energy in association with jets, as a first example.  We have derived a factorization theorem using SCET that enables the resummation of large threshold logarithms.  In our derivation we have used the global event shape N-jettiness to define the final-state jets.  N-jettiness also allows for the resummation of large logarithms associated with vetoing additional jets, further extending the usefulness of our results.  Numerical results comparing the leading-order $m_{eff}$ distribution to the next-to-leading logarithmic resummed result have been presented.  We find corrections ranging from 50\% to 100\%.  The logarithmic corrections 
increase with $m_{eff}$, potentially mimicking SUSY signatures, and should be accounted for in experimental analyses.

Several future directions remain to be pursued.  A next-to-leading order calculation of the soft function appearing in the factorization theorem would allow for an extension of the resummation accuracy to the NNLL level.  The extent to which logarithmic corrections beyond NLO affect the high-$m_{eff}$ tail could be determined.  The resummed result could then be combined with the fixed-order NLO calculation to better predict these backgrounds.  An NLO calculation of the soft function would also allow the soft scale $\mu_S$ to be determined by minimizing the contribution of the associated logarithmic terms to the cross section, as was done for the jet function.  The degree to which the threshold region is dynamically enhanced in this process could then be determined.  Our study defines jets via the N-jettiness event-shape variable, which is theoretically convenient since logarithms associated with jet vetoes can be controlled to all orders in perturbation theory.  LHC experimental studies typically define jets via the anti-$k_T$ algorithm, and it is an open question as to the quantitative effect of this difference.  Since N-jettiness jets are geometrically similar to anti-$k_T$ ones assuming the correct distance measure is chosen when defining jettiness~\cite{Jouttenus:2011wh}, we expect the difference to be small.   However, this point is worth further investigation.

In the future we plan to extend these results to include $W$ and $Z$ production in association with two and more jets, and combine the resummation with the known fixed-order results to provide a best prediction for use in experimental studies.  We also plan to study the production of other color-neutral objects in association with jets, such as the Higgs boson.  Since the Higgs is produced primarily through the gluon-gluon partonic channel, the threshold region can be enhanced even for a moderately energetic final state.  In addition, experimental Higgs searches often divide the signal into exclusive jet bins, and control over jet-veto logarithms is needed.  Our formalism handles both 
sources of logarithmic corrections, making for an interesting phenomenological application.

\section*{Acknowledgments}
We thank T. LeCompte, R. Kelley and M. Schwartz for useful discussions.  This work is supported by the U.S. Department of Energy, Division of High Energy Physics, under contract DE-AC02-06CH11357 and the grants DE-FG02-95ER40896 and DE-FG02-08ER4153, and by the U.S. National Science Foundation under grant NSF-PHY-0705682. 

\appendix
\section{RG Evolution}



Here we collect useful formulae for the evolution equations of the hard and jet functions.

\subsection{Hard Function Evolution}

As seen in Eqs.~(\ref{hard1}) and~(\ref{hard2}), the hard functions $H$ are determined via the Wilson coefficients $C$ in Eqs.~(\ref{Mtheta}) and~(\ref{MTheta}). Schematically, we can write the matching of the QCD operator ${\cal O}$ onto the SCET operators as
\bea
{\cal O} = \theta_I  C_I,
\eea 
where the indices $I$ run over the color structures shown in Eqs.~(\ref{ops1})
 and~(\ref{ops2}).  We have suppressed all spin structure. The hard function in Eqs.~(\ref{hard1}) and~(\ref{hard2}) can then be written as
 \bea
 \label{hardC}
 H_{IJ} = C_I C_J^*,
 \eea
 where we have again suppressed all spin-structure contractions on the right-hand side.

The RG evolution equation in color space for the Wilson coefficients $C_I$ is given by~\cite{Kelley:2010fn} 
\bea
\label{evo1}
\frac{\mathrm{d}C_I}{\mathrm{d}\log \mu} = \left( \,
\sum_{a \neq b}\frac{\left({\bf T}_a^A{\bf T}_b^A\right)_{IJ}}{2}\,
\Gamma_{\rm cusp}(\alpha_s) \log \frac{\mu^2}{-s_{ab}} + \,
\sum_a \gamma^a(\alpha_s)\delta_{IJ} \right) C_I \equiv \Gamma_{IJ}C_J\,,
\eea
which is valid at least up to two loops for $N$ external massless colored
particles.
Here, $s_{ab}=2\sigma_{ab}q_a \cdot q_b + i0$ with $\sigma_{ab} = 1$ if 
both $q_a$ and $q_b$ are incoming or outgoing, and $\sigma_{ab}=-1$ otherwise. 
$q_{a,b}$ denote the label momenta on the external fields on the SCET operators (see Eqs.~(\ref{ops1}) and~(\ref{ops2})), and the indices $a$ or $b$ run over all the external fields. The logarithm of $s_{ab}$ can then be written as
\bea
\log(-s_{ab}) = \log |s_{ab}| - \Delta_{ab}i \pi \,,
\eea
with $\Delta_{ab} = 1$ for $a$ and $b$ both incoming or outgoing and $0$ otherwise.  
The action of the color matrices ${\bf T}^A$  of Eq.~(\ref{evo1}) on the collinear quark and gluon fields  is defined as~\cite{Chiu:2009mg}
\bea
{\bf T}_a^A \xi_b &=& -t^A \xi_b\delta_{ab}, \nn \\ 
{\bf T}^A_a \bar{\xi}_b &=& \delta_{ab}\bar{\xi}_bt^A,\nn \\
{\bf T}^A_a A^B_b &=& \delta_{ab}A^{C}_b i f^{CAB}.
\eea
We note that in the evolution of the SCET operators, the $\Gamma_{IJ}$ on the right-hand side of Eq.~(\ref{evo1}) acts on the $\theta$ from the right: $\theta_I \Gamma_{IJ}$.
 
For the $ab\to cd+\gamma$ processes considered here, with $a,b,c,d$ denoting colored patrons, the anomalous dimension $\Gamma_{IJ}$ defined in Eq.~(\ref{evo1}) can be explicitly written as
\bea
\label{Gamma1}
\Gamma_{IJ}& = &\,
-\frac{1}{2}\left(c_1+c_2\right)\delta_{IJ}\Gamma_{\rm cusp}\,
\log\frac{\mu^2}{-s_{12}}\,
- \frac{1}{2}\left(c_3+c_4\right)\delta_{IJ}\Gamma_{\rm cusp}\log\,
 \frac{\mu^2}{-s_{34}} + \sum_a^4 \gamma^a \delta_{IJ}\nn \\
&& +\left( \frac{T_1^AT_3^A}{2}\Gamma_{\rm cusp} \left(\log\frac{-s_{12}}{-s_{13}}
+ \log \frac{-s_{34}}{-s_{13}} \right) \,
+ \frac{T_1^AT_4^A}{2}\Gamma_{\rm cusp}\left(\log \frac{-s_{12}}{-s_{14}}
+ \log \frac{-s_{34}}{-s_{14}} \right) \right.\nn\\
&& \left. + \frac{T_2^AT_3^A}{2}\Gamma_{\rm cusp}\left(\log \frac{-s_{12}}{-s_{23}}
+ \log \frac{-s_{34}}{-s_{23}}\right) \,
+ \frac{T_2^AT_4^A}{2}\Gamma_{\rm cusp} \left( \log \frac{-s_{12}}{-s_{24}}
+ \log \frac{-s_{34}}{-s_{24}} \right) \right)_{IJ} \,, \nn \\
\eea
where we have used the fact that
$\sum_aT_a = 0$ due to color-charge conservation and 
the relation $T_a \cdot T_a = c_a $, where
$c_a = C_F $ for quarks and $c_a = C_A $ for gluons.  The RG evolution equation of the hard coefficient $H_{IJ}$ can then be determined by  Eqs.~(\ref{hardC}) and~(\ref{evo1}). The  result is given by
\bea\label{Hrun}
\tilde{H}_{IJ}(\mu) &=& \tilde{H}_{IJ}(\mu_H)\,
\exp\left(2c_H S(\mu_H,\mu) - 2A_H(\mu_H,\mu)\right) \nn\\
&& \times \exp\left(\,
-A_\gamma(\mu_H,\mu)\left(\,
 \frac{c_H}{2}\log\left|\frac{(-s_{12})(-s_{34})}{\mu_H^4} \right|\,  
+2c_r\log\left|\frac{-s_{24}}{-s_{13}}\right|\,
+\lambda_I+\lambda_J^\ast \,
\right)  \right) \,,  \nn \\
\eea
where $\tilde{H}$ is related to $H$ by the transformation
\bea
\label{Hrot}
\tilde{H} = PHP^\dagger.
\eea
The transformation by the matrix $P$ corresponds to a rotation in color space that diagonalizes the matrix $\Gamma$ on the right-hand side of Eq.~(\ref{evo1}). In this basis, the evolution of the matrix elements $H_{IJ}$ in color space is multiplicative, as seen in Eq.~(\ref{Hrun}). The $\lambda_I$ in Eq.~(\ref{Hrun}) are just the eigenvalues of the matrix $\Gamma$, and  $c_r$ is a  constant. The $\lambda_I$ and $c_r$ constants depend on the partonic process. Finally, the remaining quantities in Eq.~(\ref{Hrun}) are defined as
\bea\label{SAde}
S(\mu_H,\mu)&=&-\int_{\alpha_s(\mu_H)}^{\alpha_s(\mu)}\mathrm{d}\alpha \,
\frac{\Gamma_{\rm cusp}(\alpha)}{\beta(\alpha)}\,
\int_{\alpha_s(\mu_H)}^{\alpha_s(\mu)}\frac{\mathrm{d}\alpha'}{\beta(\alpha')}\,
\hspace{2.ex}\, , \nn \\
A_{\gamma}(\mu_H,\mu) &=& -\int_{\alpha_s(\mu_H)}^{\alpha_s(\mu)}\,
\mathrm{d}\alpha \frac{\Gamma_{\rm cusp}(\alpha)}{\beta(\alpha)} \,,  \\
A_{H}(\mu_H,\mu) &=& -\int_{\alpha_s(\mu_H)}^{\alpha_s(\mu)}\,
\mathrm{d}\alpha \frac{\sum_a \gamma_{a}(\alpha)}{\beta(\alpha)} \,. \nn 
\eea
To NLL accuracy, these quantities take the form
\bea
\label{Adef}
&&S(\mu_i,\mu_f)=\frac{\Gamma_0}{4\beta_0^2}\,
\left(\frac{4\pi}{\alpha_s(\mu_i)}\left(1-\frac{1}{r} - \log r \right)
+\left(\frac{\Gamma_1}{\Gamma_0}-\frac{\beta_1}{\beta_0} \right)\,
\left(1-r+\log r \right)\,
+\frac{\beta_1}{2\beta_0}\log^2 r
\right) \nn \\
&& A_\gamma(\mu_i,\mu_f) = \frac{\Gamma_0}{2\beta_0} \left( \,
\log r \,
+\frac{\alpha_s(\mu_i)}{4\pi}
\left(\frac{\Gamma_1}{\Gamma_0}-\frac{\beta_1}{\beta_0} \right)
\left(r-1 \right)
\right) \,,
\eea 
where 
\bea
r&=&\alpha_s(\mu_f)/\alpha_s(\mu_i),\nn \\
\Gamma_0&=&4,\;\;\Gamma_1=4\left[\left(\frac{67}{9}-\frac{\pi^2}{3}\right)C_A- \frac{n_f}{9}\right],\nn \\
\beta_0 &=& \frac{11}{3}C_A-\frac{2}{3} n_f,\;\;\beta_1=\frac{34}{3}C_A^2-\left(\frac{2}{3}C_A+4C_F \right)\frac{n_f}{2},\nn \\
\gamma^a &=& \gamma_q = -3C_F\frac{\alpha_s}{4\pi} \;\;{\rm for \;quarks}, \nn \\ 
\gamma^a &=& \gamma_g = -\beta_0\frac{\alpha_s}{4\pi}\;\; {\rm  for \;gluons}.
\eea
\OMIT{
$A_H(\mu_i,\mu_f)$ can be obtained to NLL accuracy
by keeping only the $\log r$ term in $A_\gamma(\mu_i,\mu_f)$, and suitably 
replacing 
the overall coefficient via a comparison of $\sum_a\gamma_a$ with the LO 
$\Gamma_{\rm cusp}$ for each partonic channel as seen
from Eq.(\ref{SAde}).}

\OMIT{
To the leading order, $\Gamma_{\rm cusp} = 4\alpha_s/(4\pi)$ and 
$\gamma^a = \gamma_q = -3C_F\alpha_s/(4\pi)$ for quarks and 
$\gamma^a = \gamma_g = -\beta_0\alpha_s/(4\pi)$ for gluons, 
with $\beta_0 = 11/3C_A-2/3n_f$.
Also $\mathrm{d}\log\mu = 1/\beta(\alpha_s)\mathrm{d}\alpha_s $, with
$\beta(\alpha_s)= -2\alpha_s(\alpha_s/(4\pi)\beta_0 + \dots)$.
We now collect useful relations for the different types of partonic channels.}

\subsubsection{Four External Quarks}
 
 Processes with four external quarks in Eq.(\ref{qqqq}) are mediated by the operators
 \bea
 \label{ops1a}
 \theta_1 &=& \bar{\xi}_2 t^a \xi_1 \ \bar{\xi}_4 t^a\xi_3 \nn \\
 \theta_2 &=& \bar{\xi}_2 \bold{1} \xi_1\ \bar{\xi}_4\bold{1} \xi_3,
 \eea
 where we have suppressed the Dirac structure.
 With this choice of labels for the collinear fields, which is same as in Eq.~(\ref{ops1})),we have 
$T_1\cdot T_3 = T_2\cdot T_4$, $T_1\cdot T_4 = T_2\cdot T_3$ where $T_a\cdot T_b = \sum_A T_a^A T_b^A$ and
\bea
&&(T_1\cdot T_3)_{11} = -\frac{1}{C_A} \,, \,
\hspace{3.ex} (T_1\cdot T_3)_{21} = \frac{C_F}{2C_A} \,, \,
\hspace{3.ex} (T_1\cdot T_3)_{12} = 1\,,
\hspace{3.ex} (T_1\cdot T_3)_{22} = 0\,.\nn\\
&&(T_1\cdot T_4)_{11} = -\frac{C_A}{2}+\frac{1}{C_A} \,, \,
\hspace{3.ex} (T_1\cdot T_4)_{21} = -\frac{C_F}{2C_A} \,, \,
\hspace{3.ex} (T_1\cdot T_4)_{12} = -1 \,, \,
\hspace{3.ex} (T_1\cdot T_4)_{22} = 0 \,. \nn \\
\eea
Using these results in Eq.~(\ref{Gamma1}), the anomalous dimension matrix takes the form
\bea\label{dqqqq}
\Gamma_{IJ} &=& \left(\frac{1}{2}(4C_F)\Gamma_{\rm cusp} \,
\frac{1}{2}\log \frac{(-s_{12})(-s_{34})}{\mu^4} \,
+ 4\gamma_q \right)\delta_{IJ}\\
&& + \left( \frac{T_1\cdot T_3}{2}\Gamma_{\rm cusp} \,
\left(2\log\frac{(-s_{12})(-s_{34})}{(-s_{13})(-s_{24})} \right) \,
+ \frac{T_1\cdot T_4}{2}\Gamma_{\rm cusp}\,
\left(2\log\frac{(-s_{12})(-s_{34})}{(-s_{23})(-s_{14})} \right) \right)_{IJ}\,. \nn
\eea
This corresponds to the values
\bea
c_H = 4C_F \,, 
\hspace{3.ex} \sum_a\gamma^a = 4\gamma_q \,,
\hspace{3.ex} c_r = 0\,,
\eea
in Eq.~(\ref{Hrun}). The eigenvalues $\lambda_I$ of the matrix $\Gamma$ are given by
\bea
\lambda_\pm = \frac{C_A}{2} (U-T) \,
- \frac{1}{C_A} U \pm \sqrt{UT + \frac{1}{4}C_A^2(T-U)^2} \,,
\eea
where we have defined
\bea\label{UT}
U = \frac{1}{2}\log\frac{(-s_{14})(-s_{23})}{(-s_{13})(-s_{24})}\,,
\hspace{3.ex} T = \frac{1}{2}\log\frac{(-s_{12})(-s_{34})}{(-s_{13})(-s_{24})}\,.
\eea
The transformation matrix $P$ which diagonalizes $\Gamma$ is given by
\[
P  =
\left( {\begin{array}{cc}
 \lambda_+  & \frac{C_F}{C_A}U  \\
 \lambda_- &  \frac{C_F}{C_A}U \\
\end{array} } \right) \,, \hspace{4.ex} \,
P^{-1}  =
\frac{C_A}{\Delta \lambda C_F U}\left( {\begin{array}{cc}
 \frac{C_F}{C_A}U  & - \frac{C_F}{C_A}U  \\
 -\lambda_- &  \lambda_+\\
\end{array} } \right) \,.  
\]

\subsubsection{Two external quarks and gluons}

The processes with two external quarks and gluons are mediated by the operators
\bea
\label{ops2a}
\Theta_1 &=& \bar{\xi}_2 t^{a_1}t^{a_3} \xi_4A^{a_1}A^{a_3}, \nn \\
\Theta_2 &=& \bar{\xi}_2 t^{a_3}t^{a_1}\xi_4 A^{a_1}A^{a_3}, \nn \\
\Theta_3 &=& \bar{\xi}_2 \delta^{a_1a_3}\xi_4A^{a_1}A^{a_3},
\eea
where we have again suppressed the Dirac structure.
With this choice of labels for the collinear fields we  have 
$T_1\cdot T_4 = T_2\cdot T_3$ and
\bea
&&(T_1\cdot T_3)_{11} =  (T_1\cdot T_3)_{22} = -\frac{1}{2C_A} - C_F \,,
\hspace{2.ex}  (T_1\cdot T_3)_{31} = (T_1\cdot T_3)_{32}= -\frac{1}{4} \,,
\hspace{2.ex} (T_1\cdot T_3)_{33} = -C_A \, \nn \\
&&(T_1\cdot T_4)_{31} = \frac{1}{4} \,,
\hspace{2.ex} (T_1\cdot T_4)_{22} = -\frac{C_A}{2} \,,
\hspace{2.ex} (T_1\cdot T_4)_{13} = -( T_1\cdot T_4 )_{23} = 1 \,, \nn \\
&& (T_2\cdot T_4)_{11} = (T_2\cdot T_4)_{22} =\frac{1}{2C_A} \,,
\hspace{2.ex}  (T_2\cdot T_4)_{31} = (T_2\cdot T_4)_{32} = -\frac{1}{4}\,,
\hspace{2.ex} (T_2\cdot T_4)_{33} = -C_F \,.
\eea
We have only listed the non-vanishing elements. The anomalous dimension matrix is given by 
\bea\label{dggqq}
\Gamma_{IJ} &=& \left(\frac{1}{2}(2C_A+2C_F)\Gamma_{\rm cusp}\,
\frac{1}{2}\log\frac{(-s_{12})(-s_{34})}{\mu^4} \,
+ 2\gamma_q + 2\gamma_g\right)\delta_{IJ}   \\
&&+ \left(\frac{C_F-C_A}{2}\right) \Gamma_{\rm cusp}\,
\log\frac{-s_{24}}{-s_{13}}  \delta_{IJ} \nn \\
&& + \left( \left(\,
\frac{T_1\cdot T_3}{2} + \frac{T_2\cdot T_4}{2}\right)\Gamma_{\rm cusp}\,
\log\frac{(-s_{12})(-s_{34})}{(-s_{13})(-s_{24})}  \,
 +\frac{T_1\cdot T_4}{2}\Gamma_{\rm cusp} \,
\left(2\log\frac{(-s_{12})(-s_{34})}{(-s_{14})(-s_{23})} \right)\,
\right)_{IJ} \,,\nn
\eea
where we have used $\left(T_1\cdot T_3 -T_2 \cdot T_4 \right)_{IJ}
= (C_F-C_A)\delta_{IJ}$. This corresponds to the values
\bea
c_H = 2 C_A + 2 C_F \,,
\hspace{3.ex} \sum_a\gamma^a = 2\gamma_g+2\gamma_q \,,
\hspace{3.ex} c_r = \frac{1}{2}(C_F-C_A)\,,
\eea
in Eq.~(\ref{Hrun}).  The eigenvalues of $\Gamma$ and the rotation matrix $P$ can be obtained from \cite{Kelley:2010fn}. To maintain some semblance of brevity we do not provide explicit expressions for them here.
\OMIT{
And the eigenvalues $\lambda_i$ from the second line of Eq.~(\ref{dggqq})
are too lengthy but 
take the exact form as the one 
in Kelly and Schwarz's paper arXiv:1008.2759v1, 
by substituting their definitions 
of $U$ and $T$ by Eq.~(\ref{UT}). So is the transformation matrix $P$.}

\section{Tree-level expressions for the hard functions}

The hard functions $H_{IJ}$ that appear in the factorization formula of Eq.~(\ref{fac}) take the forms shown in Eqs.~(\ref{hard1}) and~(\ref{hard2}). In this section, we collect the tree-level expressions for these hard functions for the different partonic channels that contribute to $pp\to \gamma +2$ jets. The partonic channels in Eqs.~(\ref{qqqq}) and~(\ref{qqgg}) are mediated by the SCET operators in Eqs.~(\ref{ops1}) and~(\ref{ops2}), respectively.  The collinear fields in these operators are labeled by indices $\{1,2,3,4\}$. In our convention we fix the basis of operators with labels as in Eqs.~(\ref{ops1a}) and~(\ref{ops2a}). The momenta of the particles in the partonic process are chosen to correspond to a permutation of the label momenta $\{q_1,q_2,q_3,q_4\}$ of the fields. Thus, depending on the partonic process, these label momenta will correspond to a particular combination of incoming and outgoing momenta. The hard coefficients are then given in terms of these label momenta that are assigned in the partonic process. This allows for a consistency in the momentum assignments in the matrix element calculations and those that appear in the RG evolution equations. By convention, we will denote the outgoing photon momentum by $q_5$ in the following.  We note that all hard function have been checked to numerically agree with those found using Madgraph~\cite{Alwall:2011uj}.

\subsection{$ q(q_1)\bar{q}(q_4)\to Q(q_2)\bar{Q}(q_3)\gamma(q_5) $}

Here we give the hard matching coefficient for the process $ q(q_1)\bar{q}(q_4)\to Q(q_2)\bar{Q}(q_3)\gamma(q_5) $ with different quark flavors $q\neq Q$. The matching coefficient is given by
\[ H= \frac{1}{4N_c^2}(4\pi)^3\left(-4\alpha\alpha_s^2\right)\left(\frac{s_{12}^2+s_{34}^2+s_{13}^2+s_{24}^2}{s_{14}s_{23}} \right)I^2 \left( \begin{array}{cc}
\frac{1}{C_A^2} & -\frac{C_F}{C_A^2} \\
-\frac{C_F}{C_A^2} & \frac{C_F^2}{C_A^2} \end{array} \right),\]
where the four-vector $I$ is given by
\bea
I = 2\left(-e_q\frac{q_4}{s_{45}}+e_q\frac{q_1}{s_{15}}\,
-e_Q\frac{q_3}{s_{35}}+e_Q\frac{q_2}{s_{25}} \right)\,.
\eea
$e_{q,Q}$ denote the electric charges of the quarks appearing in the scattering process.

\subsection{$q(q_1)\bar{Q}(q_4)\to q(q_2)\bar{Q}(q_3)\gamma(q_5)$}

In this channel with $q\neq Q$, the hard function matrix is given by

\[ H=\frac{1}{4N_c^2}(4\pi)^3\left(-4\alpha\alpha_s^2 \right)
\left(\frac{s_{14}^2+s_{23}^2+s^2_{24}+s^2_{13}}{s_{12}s_{34}} \right)I^2\left( \begin{array}{cc}
1& 0 \\
0 &0 \end{array} \right),\]
where the four-vector  $I$ is given by
\bea
I= 2\left(-e_Q\frac{q_4}{s_{45}}+e_q\frac{q_1}{s_{15}}\,
-e_Q\frac{q_3}{s_{35}}+e_q\frac{q_2}{s_{25}} \right) \,.
\eea

\subsection{$q(q_1)\bar{q}(q_4)\to q(q_2)\bar{q}(q_3)\gamma(q_5)$}

In this channel with identical quark flavors, the hard function is given by
\bea
H= H^a + H^b + H^c,
\eea
where
\[ H^a=
\frac{1}{4N_c^2}(4\pi)^3\left(-4\alpha\alpha_s^2 \right)\,
\left(\frac{s_{14}^2+s_{23}^2+s^2_{24}+s^2_{13}}{s_{12}s_{34}} \right)I^2\,
 \left( \begin{array}{cc}
1& 0 \\
0 &0 \end{array} \right),\]
\[H^b =
\frac{1}{4N_c^2}(4\pi)^3\left(-4\alpha\alpha_s^2\right)\left(\frac{s_{12}^2+s_{34}^2+s_{13}^2+s_{24}^2}{s_{14}s_{23}} \right)I^2 \left( \begin{array}{cc}
\frac{1}{C_A^2}& -\frac{C_F}{C_A^2}  \\
-\frac{C_F}{C_A^2}  &\frac{C_F^2}{C_A^2} \end{array} \right),\]
\[H^c= 
\frac{1}{4N_c^2}(4\pi)^3 \left(-4\alpha\alpha_s^2\right) \,
\left(s^2_{13}+s^2_{24}\right)
\left(\frac{s_{13}s_{24}-s_{12}s_{34}-s_{14}s_{23}}{s_{14}s_{23}s_{12}s_{34}} \right)\,
I^2 \left( \begin{array}{cc}
-\frac{1}{C_A}& \frac{C_F}{2 C_A}  \\
\frac{C_F}{2 C_A}  &0\end{array} \right).\]
The four-vector $I$ is given by
\bea
I= 2\left(-e_q\frac{q_4}{s_{45}}+e_q\frac{q_1}{s_{15}}\,
-e_q\frac{q_3}{s_{35}}+e_q\frac{q_2}{s_{25}} \right) \,.
\eea

\subsection{$q(q_1)Q(q_3)\to q(q_2)Q(q_4)\gamma(q_5)$}

In this channel for $q\neq Q$ the hard function is given by
\[
H = \frac{1}{4N_c^2}(4\pi)^3 \left(-4\alpha\alpha_s^2 \right)\,
\left(\,
\frac{s_{13}^2+s_{24}^2+s_{14}^2+s_{23}^2}{s_{12}s_{34}} \,
\right)I^2  \left( \begin{array}{cc}
1& 0 \\
0 &0 \end{array} \right),\] 
where the four-vector $I$ is given by
\bea
I = 2\left( \,
e_{q}\frac{q_1}{s_{15}}+e_{Q}\frac{q_3}{s_{35}}\,
+e_{q}\frac{q_2}{s_{25}}+e_{Q}\frac{q_4}{s_{45}}\,
\right)\,.
\eea

\subsection{$q(q_1)q(q_3)\to q(q_2) q(q_4)\gamma(q_5)$}

In this channel with identical quarks the hard function is given by
\bea
H= H^a+H^b+H^c,
\eea
where
\[
H^a = \frac{1}{4N_c^2}(4\pi)^3 \left(-4\alpha\alpha_s^2 \right)\,
\left(\,
\frac{s_{13}^2+s_{24}^2+s_{14}^2+s_{23}^2}{s_{12}s_{34}} \,
\right)I^2 \left( \begin{array}{cc}
1& 0 \\
0 &0 \end{array} \right),\] 
and
\[H^b =  \frac{1}{4N_c^2}(4\pi)^3  \left(-4\alpha\alpha_s^2 \right)\left(\frac{s_{12}^2+s_{34}^2+s_{13}^2+s_{24}^2}{s_{14}s_{23}} \right)I^2\left( \begin{array}{cc}
\frac{1}{C_A^2}& -\frac{C_F}{C_A^2} \\
 -\frac{C_F}{C_A^2} &\frac{C_F^2}{C_A^2} \end{array} \right),\] 
\[ H^c =   \frac{1}{4N_c^2}(4\pi)^3  \left(-4\alpha\alpha_s^2 \right)\,
(s_{13}^2+s_{24}^2) \left(\,
\frac{s_{13}s_{24}-s_{12}s_{34}-s_{14}s_{23}}{s_{14}s_{23}s_{12}s_{34}} \,
\right)I^2 \left( \begin{array}{cc}
-\frac{1}{C_A}& \frac{C_F}{2 C_A} \\
 \frac{C_F}{2 C_A} & 0 \end{array} \right).\] 
The four-vector $I$ is given by
\bea
I = 2\left( \,
e_{q}\frac{q_1}{s_{15}}+e_{q}\frac{q_3}{s_{35}}\,
+e_{q}\frac{q_2}{s_{25}}+e_{q}\frac{q_4}{s_{45}}\,
\right)\,.
\eea

\subsection{ $g(q_1,a_1)g(q_3,a_3)\to q(q_2)\bar{q}(q_4) \gamma(q_5)$}

For this gluon-initiated process, we find the hard function 

\[  H= \frac{1}{4(N_c^2-1)^2}\left((4\pi)^3\alpha \alpha_s^2\right)16 e_q^2 \,
\frac{\sum_{i=1,3,5}s_{i2}s_{i4}\,
\left(s_{i2}^2+s_{i4}^2 \right)}{s_{13}\prod_{i=1,3,5}s_{i2}s_{i4} }\\ \]
\[ \times 
\left( \begin{array}{ccc}
s_{14}s_{23} & \frac{1}{2}(s_{13}s_{24}-s_{14}s_{23}-s_{12}s_{34} )& 0\\
\frac{1}{2}(s_{13}s_{24}-s_{14}s_{23}-s_{12}s_{34} )& s_{12}s_{34} & 0  \\
0&0&0 \end{array} \right) . \]

\subsection{$q(q_4){\bar q}(q_2) \to g(q_1,a_1) g(q_3,a_3) \gamma(q_5)$}

For this scattering process we find the hard function

\[  H= \frac{1}{4N_c^2}\left((4\pi)^3\alpha \alpha_s^2\right)16 e_q^2 \,
\frac{\sum_{i=1,3,5}s_{i2}s_{i4}\,
\left(s_{i2}^2+s_{i4}^2 \right)}{s_{13}\prod_{i=1,3,5}s_{i2}s_{i4} }\\ \]
\[ \times 
\left( \begin{array}{ccc}
s_{14}s_{23} & \frac{1}{2}(s_{13}s_{24}-s_{14}s_{23}-s_{12}s_{34} )& 0\\
\frac{1}{2}(s_{13}s_{24}-s_{14}s_{23}-s_{12}s_{34} )& s_{12}s_{34} & 0  \\
0&0&0 \end{array} \right) . \]

\subsection{$q(q_4)g(q_1)\to q(q_2)g(q_3)\gamma(q_5)$}

For the $qg$ channel we obtain the hard function
\[  H=  \frac{1}{4N_c(N_c^2-1)}\left(-(4\pi)^3\alpha \alpha_s^2\right)16 e_q^2 \,
\frac{\sum_{i=1,3,5}s_{i2}s_{i4}\,
\left(s_{i2}^2+s_{i4}^2 \right)}{s_{13}\prod_{i=1,3,5}s_{i2}s_{i4} }\\ \]
\[ \times 
\left( \begin{array}{ccc}
s_{14}s_{23} & \frac{1}{2}(s_{13}s_{24}-s_{14}s_{23}-s_{12}s_{34} )& 0\\
\frac{1}{2}(s_{13}s_{24}-s_{14}s_{23}-s_{12}s_{34} )& s_{12}s_{34} & 0  \\
0&0&0 \end{array} \right) . \]
It takes exactly the same form for other $qg$ channels, such as
${\bar q}(2)g(1) \to {\bar q}(4) g(3) \gamma(5)$.



\section{SCET jet and soft functions}
\label{jet}
In this section, we present the definitions of both the jet functions and
the soft function used in our approach. The jet functions defined in
the framework of SCET have been known for some time up to the two-loop order~\cite{Bauer:2010vu,Fleming:2003pr,Becher:2006pl,Becher:2011pl}. We refer the readers to the operator definition
of the jet functions therein and we note that the jet functions $J(s)$
used in our current work are normalized to $\delta(s)$ at LO.
The soft function showed here characterizing the 
restricted soft radiation inside and outside the jet region 
near threshold is new.

The operator defintion of the soft function is
\bea
S_{JI}\left(k_{out}^0,\tau_{N,s}^{J_i}\right) = \,
\langle 0 | \,
{\cal O}_J^{s\dagger}\,
\delta\left(k_{out}^0 - \hat{k}^0\right)
\prod_{J_i} \delta\left(\tau_{N,s}^{J_i} - \hat{\tau}_{N}^{J_i} \right)
{\cal O}_I^s\,
|0 \rangle  \,,
\eea
where ${\hat k}^0$ and ${\hat \tau}_N^{J_i}$ are the operators
which act on the final ultrasoft states to project out the related observables.
${\cal O}^s$ is a collective of ultrasoft Wilson lines $Y$ and 
color structures ${\bf T}$ 
appearing in the SCET operators in Eqs.~(\ref{ops1}) and~(\ref{ops2}), which formally can be written as
\bea
{\cal O}^s = T\left[Y^\dagger_i {\bf T} Y_j Y^\dagger_k {\bf T}Y_l \right]\,.
\eea 
The subscripts of $Y$'s correspond to the subscripts of
the fields in each SCET operator.

In this manuscript, the jet functions are run from the jet scale 
to the soft scale via the RG evolution equation
\bea
\frac{\mathrm{d}J_a(s,\mu)}{\mathrm{d}\log\mu} = \int ds' \,
\left[-2C_a\Gamma_{\rm cusp} \frac{1}{\mu^2}\,
\left(\frac{\mu^2}{s-s'} \right)_+ +\gamma_J^a \delta(s-s') \right]\,
 J_a(s',\mu)\,.
\eea
The index $a$ runs over $\{q,g\}$ corresponding to quark and gluon jet functions respectively, 
 $C_q = C_F$ and $C_g = C_A$. To leading order,
\be
\gamma_J^q = 6C_F\frac{\alpha_s}{4\pi},\;\; \gamma_J^g = 2\beta_0\frac{\alpha_s}{4\pi}.
\ee 
In position space with $F(y) = \int\mathrm{d}s \exp(-is y)F(s) $, this equation becomes
\bea
\frac{\mathrm{d}J_a(y)}{\mathrm{d}\log\mu} = \,
\left(2C_a\Gamma_{\rm cusp} \log(i y \mu^2 e^{\gamma_E}) + \gamma^a_J \right)\,
J_a(y)\,.
\eea
The solution is
\bea
J_a(y,\mu) = \exp(-4C_a S(\mu_J,\mu) - A_J^a(\mu_J,\mu))\,
\left(iy\mu_J^2e^{\gamma_E} \right)^{-2C_aA_\gamma(\mu_J,\mu)} J_a(y,\mu_J)\,.
\eea
Here $A_J^a$ is obtained by replacing $\Gamma_{\rm cusp}$ by $\gamma_J^a$ 
in $A_\gamma$ defined in Eq.~(\ref{Adef}).  For the product of jet functions that appears in the factorization theorem, we have 
\bea
J_aJ_b = \exp(-4(C_a+C_b) S - A_J^a-A_J^b)\,
\left(iy\mu_J^2e^{\gamma_E} \right)^{-2(C_a+C_b)A_\gamma} J_a(y,\mu_J)J_b(y,\mu_J)\,.
\label{jmult}
\eea
 
The NLO jet functions are used to determine the jet scale
in our current work and are given by
\bea
&&J_q(s)=\delta(s)+\frac{\alpha_sC_F}{4\pi}\,
\left((7-\pi^2)\delta(s)-\frac{3}{\mu^2}{\cal L}_0(s/\mu^2)\,
+\frac{4}{\mu^2}{\cal L}_1(s/\mu^2) \right)\,, \nn \\
&&J_g(s)=\delta(s) + \frac{\alpha_s}{4\pi}\,
\left(\left(\left(\frac{4}{3}-\pi^2\right)C_A + \frac{5}{3}\beta_0\right)\,
\delta(s)-\frac{\beta_0}{\mu^2}{\cal L}_0(s/\mu^2) \,
+\frac{4C_A}{\mu^2}{\cal L}_1(s/\mu^2)\, 
\right)\,.\nn \\
\eea

\OMIT{
Last we quote the useful relation, under fourier transformation
\bea
(iy)^\w \to \frac{1}{\Gamma(-\w)}\,
\left(\frac{\theta(s)}{s^{1+\w}} \right)_+ \,
\to^{\w\to 0} \delta(s)
\eea}
For NLL accuracy we need only the LO soft function which takes the form
\bea
S^{(0)}_{JI} = {\bf S}_{JI}\,
 \delta\left(k_{out}^0\right) \,
\prod_{J_i} \delta\left(\tau_{N,s}^{J_i}\right)\,,
\eea
where for the four-quark channels the color matrix ${\bf S}$ is diagonal and reads

\[ {\bf S} = \left ( \begin{array}{cc} 
\frac{C_A C_F}{2} &2 \\
0& C_A^2
\end{array} \right ), \]
while for the two-quarks and two-gluon configurations the color
matrix ${\bf S}$ is 
\[ {\bf S} = \left ( \begin{array}{ccc} 
C_A C_F^2& -\frac{1}{2}C_F & C_A C_F \\
-\frac{1}{2}C_F & C_A C_F^2 & C_A C_F \\
C_A C_F & C_A C_F &  2C_F C_A^2
\end{array} \right ). \]

\end{document}